# Appropriate Methodology of Statistical Tests According to Prior Probability and Required Objectivity

**Tomokazu Konishi**[1,*]

[1] Faculty of Bioresource Sciences, Akita Prefectural University, Shimoshinjo Akita 010-0195, JAPAN

[*]email: konishi@akita-pu.ac.jp

SUMMARY.  In contrast to its common definition and calculation, interpretation of *p*-values diverges among statisticians. Since *p*-value is the basis of various methodologies, this divergence has led to a variety of test methodologies and evaluations of test results. This chaotic situation has complicated the application of tests and decision processes. Here, the origin of the divergence is found in the prior probability of the test. Effects of difference in Pr($H_0$ = true) on the character of *p*-values are investigated by comparing real microarray data and its artificial imitations as subjects of Student's *t*-tests. Also, the importance of the prior probability is discussed in terms of the applicability of Bayesian approaches. Suitable methodology is found in accordance with the prior probability and purpose of the test.

**1 Introduction**

Statistical tests are frequently used to objectively evaluate whether an observation is significant, and the *p*-value provides essential information. Indeed, testing and *p*-values may have become so popular that indiscriminate usages are criticized (Cohen, 1994; Nelder, 1999). In spite of their popularity, the physical interpretation or rationale of a *p*-value has diverged among statisticians (Kempthore, 1976; Cox, 1977; Lehmann, 1993; Cohen, 1994; Royall, 1997; Berger, 2003; Hubbard et al, 2003; Pereira et al, 2008) while its mathematical definition has been commonly recognized as *p*-value = Pr($T \geq t_{obs}$ | $H_0$ = true), where $t_{obs}$ is the observed value of a test statistic *T*. The divergence of rationale is obvious among three major schools of statistics: Fisher, Neyman-Pearson, and Bayesian (Cox, 1977; Berger, 2003). It clearly appears in the supportive character of *p*-value: Fisher (Fisher, 1928) insisted that "a smaller *p*-value indicates stronger evidence," but Neyman-Pearson (Neyman and Pearson, 1933) did not take the *p*-value as evidence and instead stuck to the value of the threshold. Additionally, Bayesian procedures estimate posterior probability, and Bayesians may contest the validity of *p*-value (Lindley, 1957; Cohen, 1994). Indeed, the different rationales can lead to quite different conclusions, because many "positive" *p*-values would be much smaller than the threshold. the differences actually calculated can become large enough to amaze us. Also, p-value and posterior probability can become quite different; although they are not identical in a mathematical sense, actual calculated differences can become amazingly, even shockingly, large, as was shown by Lindley (1957).

The diversity of interpretations reflects the diversity of philosophies of statistical testing. As physical interpretation explains what *p*-value means in the context of the cases being tested, it clarifies how a mathematical model is applied to the data in a test methodology. Which model is applied and how it is applied to the particular data is a philosophical issue, and methodologies are coherent within the philosophy that they are based on. Because of the fundamental divergence, different methodologies will give different answers to identical observations. Hence, the divergence has created difficulties in all aspects of testing, from choosing appropriate methodology to evaluating the results, and has weakened the tests' objectivity. Furthermore, the discrepancy could reflect the fragility of the basis that we rely on to observe, evaluate, and decide.

However, it is still possible that appropriate philosophy, and thus interpretation of *p*-value, is different for different cases. Each case has its own background, such as population characteristics and production mechanisms, sample selection and measurement, and sources of data fluctuations. A system or structure can be supposed to organize the background, although the total image of the structure is rarely given. The prior probability, $\Pr(H_0 = \text{true})$, could be derived from the underlying system of data, since $H_0$ is often settled on some characteristics of the populations. The prior probability would alter the character of the *p*-value as a random variable; indeed, Hung et al. (1997) noted that the distribution is known to be uniform under the null hypothesis, and showed that it will be skewed when $H_0$ is false. This alteration of the distribution will influence how *p*-values should be interpreted. Additionally, the prior probability is critically important in Bayesian procedures as well. Hence, the prior probability could alter conclusions obtained by any of the schools' test methodologies.

Here, effects of the differences in the prior probability on *p*-value are observed in some extreme examples: expressional microarray data and its artificial imitations. To observe the character of *p*-values the most desirable data is the repetition of tests on a constant condition, but such data is practically unobtainable. Microarray data was chosen instead, since the methodology measures a huge number of items of the same category and gene expression levels; the items may have some variations among the conditions, such as stability of expression levels, but they share the underlying system. Therefore, by using the large number of items, characteristics of *p*-values could be observed well. For example, if *p*-values for each of the genes are uniformly distributed, *p*-values for the data set will also become uniform. Likewise, if *p*-values support evidence, those obtained in related experiments will take similar values for each gene, which could be found in correlations among the experiments.

Microarray data are frequently subjected to statistical tests. Experiments are often repeated to manage sample variations, forming a data matrix with a large number of "gene" items and some groups of "biological conditions" that consists of several repeated experimental samples. In each gene, we use Student's *t*-test or analysis of variance, taking $H_0$ to mean that population means are identical among the groups, to test significance in expressional changes among groups. Here, we used data obtained in time-course experiments of cell differentiation (Hailesellasse Sene et al, 2007), with n=3 experimental replicates; each time point was compared with the control (0h), and the distributions of *p*-values for the genes were investigated (Figure 1A). The differentiation process was investigated separately in three lines of stem cells.

In testing such microarray data, $H_0$ is rarely true for two reasons. The first reason is in the cellular mechanisms that control expression levels of genes. Functional molecules that form the mechanisms have been well studied, and their functions are inclusively elucidated by laws of thermodynamics (Konishi, 2005). Thus, each of the cellular factors recognizes specific nucleotide sequences of genomic DNA or mRNA species, and alters expression levels of those genes. Any conditional alterations would affect several factors, changing expression levels of many genes that primarily respond to the alteration. Then, other genes will be affected somehow to follow up alterations caused by the primary genes; indeed, genes have complex relationships that are frequently compared to a network. Another reason for the low $Pr(H_0 = \text{true})$ is that the $H_0$ is a sharp or point null hypothesis; i.e., a two-sided hypothesis testing a continuous random variable by specifying the value. Although microarray data are digital, they are made by the sum of numbers of pixel data, hence they are continuous in a practical sense. Therefore, probability that a population mean takes any particular value will be null, thus complete coincidence of population means would be rare.

For comparison, some series of data that have high $Pr(H_0 = true)$ were prepared. One series used the same cell differentiation data, but *t*-tests were performed in artificial combinations of data; three data groups were split and reunited to form three groups (Figure 1B). The new groups share population; i.e., mix the three cell lines. One other series is artificial data, which mimics the differentiation data by using random numbers (Materials and Methods). Having the same variance and mean observed in each gene of the original data, they are similar to microarray data, and have certain variety that mimics noise; however, the between-group differences were removed. The condition may correspond to comparisons among samples taken from a population; real data in the condition were found in mock controls of microarray experiments that observed effects of dioxin on the kidney (Boutros et al, 2009). Among the three series, all $H_0$ are true although they are sharp; however, the balance of within- and between-group variances diverges.

## 2 Materials and Methods

The microarray data used were obtained from the public database NCBI GEO (Barrett et al, 2009). The time course of cell differentiation experiments (Hailesellasse Sene et al 2007), under series accession of GSE2972, GSE3749, and GSE3231 from each cell line, were measured on Mouse Genome 430A 2.0 Array, while the dioxin responses (Boutros et al 2009), under id GSE15859, used Mouse Genome 430 2.0 Array (Affymetrix Inc., CA). The ".cel file" data were normalized using SuperNORM data services of Skylight-Biotech Inc. (Akita); the perfect match data were parametrically normalized to obtain *z*-scores, and summarized to estimate the expression level of each gene by taking the trimmed means (Konishi, 2008). Both the normalized .cel files and the summarized levels are deposited in NCBI GEO under accession of GSE26498. In this article, only the summarized levels were used for simplifying the calculations.

In the time course, a group of repeated experiments for a time point was compared to the control group, 0h, at each gene's summarized level of a cell line (Figure 1A), and the *p*-value was calculated by applying Student's *t*-test between the groups. Positive genes were found by using a threshold of 1% in a two-sided test, 0.005. The level of noise, which may be provoked by differences among cell cultures, was estimated by using the arithmetic mean of within-group variances found at each gene. Likewise, the level of signal, which represents the expressional changes, was measured from means of the group's data, *x*, at 0h and time point *t* at gene *g* as

$\sum_g n_r (\bar{x}_{t,g} - \bar{x}_{0h,g})^2 / 2n_c$, where $n_r$ and $n_c$ are the number of repeated measurements and gene contents, respectively.

To simulate high $\Pr(H_0 = \text{true})$ cases, the microarray data was mimicked by using normally distributed random numbers, producing the same dimension of artificial data. The data were generated in a gene-wise manner, using population standard deviations and population means identical to those estimated in the corresponding gene of the real microarray data of the cell lines. The p-values were calculated in the same manner as for the real data. To simulate low $\Pr(H_0 = \text{true})$ cases, a constant, 0.6 times the standard deviation, was added to one of the groups' data in each gene. This gave the level of signal, 0.0325, the average of between-group variances, a likely level at 6-18h of the real microarray data. Additionally, to emulate high $\Pr(H_0 = \text{true})$ cases by using real microarray data, two series of artificial combinations of data were prepared, as explained in Figure 1B and 1C.

Observed values were compared between the corresponding genes in scatter plots. The expressional changes were estimated in $\Delta z$; i.e., differences between the summarized scores, which are linear to log ratios (Konishi, 2008). Spearman's nonparametric method was applied to indicate correlations, as fluctuations of p-values may not be normally distributed, and as taking logarithms of p-value does not alter the rank-based method.

The binominal test for coincidence of selected genes was performed under $H_0$ so that the p-value-based selections of genes do not have correlations among cell lines. The hypothesis is satisfied if p-values are derived by random effects. Under $H_0$, a gene will be selected in both cell lines coincidentally at a probability of (230/22,626) times (308/22,626). The expected numbers of selected genes will obey binominal distribution, with the presented probability and 22,626 trials. The probability expected for 13 or more genes is selected common to both cell lines and is estimated for the p-value.

All calculations were performed using the R (R Development Core Team, 2007); scripts for reproducing the figures from the deposited data are available in the data supplement.

**3 Results**

To observe the differences given by $\Pr(H_0 = \text{true})$, distributions of p-values were compared among the real and artificial data (Figure 2). All the histograms obtained in the time-course experiments were tilted or skewed (those for 6h and 14d are presented in Panels A and B, respectively). The degrees of skew were different among the time points. The average of between-group variances increased throughout the time course, while within-group variances were rather stable, showing the expansion of expressional changes and steady noise levels (Panel C). In contrast to the microarray data, in the simulation of high $Pr(H_0 = true)$ cases in the artificial data, the distribution of *p*-values was uniform (Panel D, gray bars). The distribution skewed when certain between-group differences were given to the artificial data (Panel D, black line); this alteration changed $Pr(H_0 = true)$ to be null. In another example of true nulls, rearranging groups (Figure 1B), higher *p*-values were observed at higher frequency (Figure 2E), unlike the expectation of uniform distribution (Hung et al 1997). In comparisons between subgroups artificially settled within a group (Figure 1C), the distributions were almost uniform (Figure 2F, gray bars); however, in some alternative combinations of data, the distributions were skewed (black line); degree and direction of the skew depended on data combinations (not shown).

When the null hypothesis is true, the *p*-value would be determined by noise; hence the value will not support evidence. On the other hand, when the *p*-value supports evidence, those found in likely conditions will show correlations. To check the possibilities, *p*-values were compared in high and low $Pr(H_0 = true)$ cases. Observed *p*-values showed different levels of correlations (Figures 3A-3C); under true null hypotheses, no correlation was observed between corresponding tests (Panel A), while the degree of correlation increased as the time course proceeded (cases in 6h and 14d are presented in Panels B and C, respectively); the oblique extension was reinforced and the Spearman's correlation rho increased (appeared in the panels). Similar tendencies were also observed among alternative combinations with another cell line, and *p*-values did not show correlations among the other high $Pr(H_0 = true)$ cases (not shown).

At 6h in the time course, total expressional differences did not show clear correlations between cell lines (Figure 3D). Actually, the number of positive genes, i.e., genes with *p*-values equal or less than 0.005, was rather small: 230 and 308 genes for experiment GSE2972 and GSE3231, respectively, out of 22,626 gene contents (a histogram for one of the cell lines is in Figure 2A). Also, the average of between-group variances was smaller than those at later time points (Figure 2**C**). Among the positive results, 13 genes were found commonly in the cell lines; this is significantly large (p = 2.49e-5) in a binominal test (Materials and Methods), showing correlation between the selections performed in the two cell lines. Additionally, in the commonly selected genes, expressional changes coincided (Figure 3E). In the later time point at 14d, 1,273 positive genes showed obvious correlations between the cell lines (Figure 3F). Such correlations among genes with low *p*-values were also observed in alternative pairs with another cell line and not in any of the artificial data (not shown). These observations are reasonable since they suggest a similar process of differentiation among cell lines and the effect of noise in the artificial data.

# 4 Discussion

Among the examples observed here, the supportive character of the *p*-value diverged according to $\Pr(H_0 = \text{true})$. When the prior probability was low, *p*-values showed weak correlations among similar experiments (Figures 3B and 3C), and expressional changes coincided between genes with smaller *p*-values (Figure 3E and 3F). Hence, under this condition, the *p*-value supports evidence. The opposite character of the *p*-value was observed when the prior probability was high. Consequently, neither Fisher's nor Neyman-Pearson's physical interpretations of *p*-value are right or wrong; rather, they are appropriate only in specific cases. If $H_0$ is highly probable, Neyman-Pearson's interpretation is appropriate, while Fisher's interpretation is appropriate when $H_0$ is unlikely. Additionally, it should be noted that the prior probability is an aspect of the underlying system of data, and other aspects could alter *p*-values as well.

When null hypotheses were certainly false, the distributions of *p*-values skewed as predicted (Hung et al 1997), and the degrees of skew varied (Figures 2A and 2B). The variation seemed to have derived from increasing between-group variances (Figure 2C), which can be explained by two alternative scenarios: the number of responding genes increased, or the magnitude of responses increased, as the time-course experiment proceeded. Both scenarios sound reasonable, although they are different in the $\Pr(H_0 = \text{true})$. However, the tilted distribution in Figure 2A cannot be explained by a mixture of a uniform and a heavily skewed distribution; this negates the high $\Pr(H_0 = \text{true})$ scenario. Rather, the tilted distribution could be mimicked by certain levels of between-group differences introduced into the artificial data (Figure 2D, black line), suggesting that the shape of the histogram in Figure 2A, the slope and the flat region, may have been formed by weaker evidence that could not reject $H_0$. Therefore, the prior probability might be kept low even in the early stages of cell differentiation.

Contrary to the prediction of uniform character (Hung et al 1997), the real distribution under the null hypothesis skewed (Figures 2E and 2F). True null hypotheses were not a sufficient condition for the distribution; rather, a balance of between- and within-group variances is necessary. When both of the variances are determined by the same level of noise, as is the case in the artificial imitation of microarray data, the condition will be satisfied as is obvious in Figure 2D; however, trivial differences can affect the distribution, such as in the skew found in Figure 2F.

Although histograms of *p*-values are valuable and rarely provided, estimation of $\Pr(H_0 = \text{true})$ is difficult to achieve based on such simple observations. Indeed, in Figure 2, shapes of the histograms diverged regardless of $\Pr(H_0 = \text{true})$, negating the appropriateness of such attempts. For example, in a methodology introduced for microarray data, the prior probability is estimated by extrapolating the flat region of the histograms (Storey and Tibshirani, 2003). It is true that such flat regions frequently appear in the histograms (like the cases in Figure 2A and 2B); however, the flat region may not reflect the expected number of true nulls. Indeed, many, if not all, of the higher *p*-values can be explained by weaker evidence, as was discussed. Additionally, the estimation of the probability will become larger than one in some extreme cases, such as in Figure 2E.

To maintain the objectivity of the test, the prior probability should be estimated in a comprehensive manner by knowing the underlying system of data. As observed in the examples presented, some of the very high or low $\Pr(H_0 = \text{true})$ cases are obvious from the background of the data. Fortunately, as the number of observations increases, the accuracy of estimation will be improved in the other cases. For example, among cases for which the underlying system of data is given, $\Pr(H_0 = \text{true})$ for each probe of microarray-based study for single nucleotide polymorphisms (SNPs) may vary with races or roots of the sample. Since an observation could determine therapeutic strategy, information on the credibility of the observation is desirable, so a Bayesian approach would be useful. Indeed, generic prior probability is estimated within a chip data and used to judge genotype calling for any probes (Affymetrix, 2006); however, as the number of samples increases, frequency data for particular SNPs for particular races or roots will accumulate, and we can feed back the frequency to the estimation of prior probability for each combination of probe and sample. Knowledge of the underlying system of data enables such attempts. Because the prior probability will affect the appropriateness of a methodology, it should be estimated by using a suitable model and evidence; otherwise, we cannot ensure the objectivity of the test. Indeed, applying statistical methods without understanding underlying systems of data will cause problems (Diaconis, 1985; Cohen, 1994; Nelder, 1999); one of the reasons for this would be the lack of suitable information to estimate the prior probability. Actually, insufficient awareness of diversity in underlying systems of data may have influenced the philosophy of each statistician through familiarity with data, producing the persistence of certain physical interpretations.

We can reevaluate the appropriateness of some methodologies in view of the prior probability. For example, adjustment for multiplicity of tests in simultaneous inference would be inappropriate in some cases. As the number of tests increases, expectations of the number of false positives may increase. However, ways of coping with the problem diverged greatly among statisticians, from "nonmultiple comparisonists" to "ultraconservative statisticians" (Miller, 1981).. Under a high $Pr(H_0 = true)$ condition, any positive results could be false, therefore multiple tests inevitably increase errors. To overcome this problem, adjustments were very conservative. Indeed, most, if not all, adjustments follow the worst scenario, $Pr(H_0 = true) = 1$ (Rothman, 1990). On the contrary, when $H_0$ is false, Type I error will not appear. Therefore, if an adjustment is applied to data with a low $Pr(H_0 = true)$, the probability of a Type II error, a false negative, will become unnecessarily huge, masking the observations. Actually, if we perform a Bonferroni adjustment to keep a family-wise error rate for the chip contents, no genes at 6h of the time-course experiment turn positive in any of the cell lines; hence, no information could be extracted as in Figure 3E.

The distribution of *p*-values would indicate some problems. For example, tendency toward a higher *p*-value in the case of Figure 2E indicates dominance of within-group variance. Although the example was obtained under the null hypothesis, this type of skew could occur in low $Pr(H_0 = true)$ cases also, from technical problems occurring independent of the groups, such as random infections in test animals. If we can find and solve the problem, the sensitivity of measurement can be improved in the next experiment. However, smaller *p*-values still indicate positive results even in such cases; at the least, the skewed distribution suggests that noise affected between- and within-group variances independently. On the contrary, if the distribution is uniform, we should consider the possibility that random effects have determined every part of the data. Indeed, in extreme cases, noise can completely hide signals; in such cases, *p*-values will not support evidence regardless of the prior probability. Unfortunately, we cannot discover false positive errors caused by wrongly added between-group variances; such errors alter the distribution concentrated in lower values, and this type of skew is indistinguishable from true positive results. Such inflation of between-group variance can be derived by inclined samplings, poor experimental designs, or infections occurring in a cage.

Although *p*-values support evidence in low $Pr(H_0 = true)$ cases, the value would not reflect signal but signal-to-noise ratio, and hence the value will be sensitive to noise. Indeed, the correlations among different cell lines were weaker than those found in the signals (compare Figure 3C and 3F). Evidence supported by *p*-value is not the magnitude but the reliability of observation.

As does the interpretation of *p*-values, the purpose of statistical tests diverges. The diversion itself is not problematic; problems occur only if there is a conflict between the methodology and the purpose. For example, as with the microarray data presented here, the necessity of the tests for very low Pr(H$_0$ = true) cases may seem to be obscure, since the null hypothesis is false anyway. In practice, however, results of tests can end up negative, as can be seen in Figure 2A. Here, *p*-values give information to judge whether there is sufficient evidence to reject H$_0$. Actually, in the earlier time points of cell differentiation, between-group variances were relatively small and many of them could be hidden by noise (Figures 2C and 3D). However, the coincidence observed among genes with lower *p*-values (Figure 3E) suggests that we can rely on the positive results. Thus, the tests have enabled us to extract meaningful information from a relatively high level of noise, by using a totally objective method. Additionally, we can rely on not only the genes commonly selected among different cell lines but all the positive results. In genes that were positive in only one of the experiments, the correlations could not be checked just because information for their counterparts was missing. Hence, there is no reason to exclude the positive observations in each of the experiments. It is natural that a smaller number of genes will be selected repeatedly by escaping from noise. Indeed, only two of them were selected commonly among the three cell lines.

In Bayesian methods, *p*-value itself will become less important, and interpretation of the posterior probability does not diverge as in the case of *p*-values. Additionally, many Bayesian models accommodate for the multiplicity, making further adjustments unnecessary (Berger, 2006). However, Bayesian methodologies have limits to their applicability. For example, they are difficult, if not impossible, to apply to a boundary condition of prior probability, such as in sharp hypotheses (Berger and Sellke, 1987; Pereira et al 2008) . Also, prior probabilities are rarely given with objectivity in the calculation. There seem to be two alternative Bayesian solutions for cases without sufficient information for prior probability: subjective (Goldstein, 2006)  and objective (Berger, 2006) approaches. However, objectivity in the objective approach may have certain limitations, since although varieties of models are being used (Berger, 2006), their appropriateness is difficult to confirm (Fienberg, 2006). Hence, the applicability varies according to how objective the test should be, and this depends on the purpose of the test. In some cases, we have to make an optimum decision on the existing knowledge; in these cases, objectivity could be restricted. In other cases, we should postpone the decision, aware that there is not enough evidence. This would be true of many cases in experimental science, since required studies could be appended afterwards. Rather, objectivity is rigorously important when integrating observations contributed by many researchers, and observations with insufficient evidence could cause unnecessary confusion. Therefore, the applicability of models with limited objectivity should be considered on the balance of benefit to risk.

# 5 Conclusion

Whether a *p*-value supports evidence or not depends on the test's prior probability. Therefore, physical interpretation of *p*-value diverges among cases, and appropriate methodologies should be found accordingly. The prior probability should be estimated in a comprehensive manner; feedback processes can improve the accuracy of the estimation. When sufficient information is not available for estimating the prior probability, appropriate methodologies are determined by requirements for the test's objectivity, based on the test's purpose. Therefore, currently recommended test methodologies can be found, as summarized in Table 1. The first branching point is in the prior probability of the test. If it is not given with objectivity, the next branching point is in the purpose of the test: "Should something be decided with the present knowledge?" Then suitable methodology will be determined, leading to indicators of the reliability of observation. Philosophy might have variations among statisticians and hence could be vague. Here, the definitions are found in the supportive character of *p*-values and use of posterior probabilities. The term "basic Bayesian" represents approaches using extra data, such as could be seen in the feedback process of SNPs.

**Table 1.** Summary for the recommended philosophy for test methodologies

| Pr(H0 = true) | Decision urgency | Recommendation | Indicator obtained | Multiplicity adjustment |
|---|---|---|---|---|
| quite low | - | Fisher | *p*-value | not required |
| given | - | Bayesian (basic) | posterior probability | |
| not given | yes | Bayesian (subjective) | | |
| not given | yes/no | Bayesian (objective) | | |
| not given | no | suspend | - | - |
| possibly high | - | Neyman-Pearson | threshold | required |

FIGURES

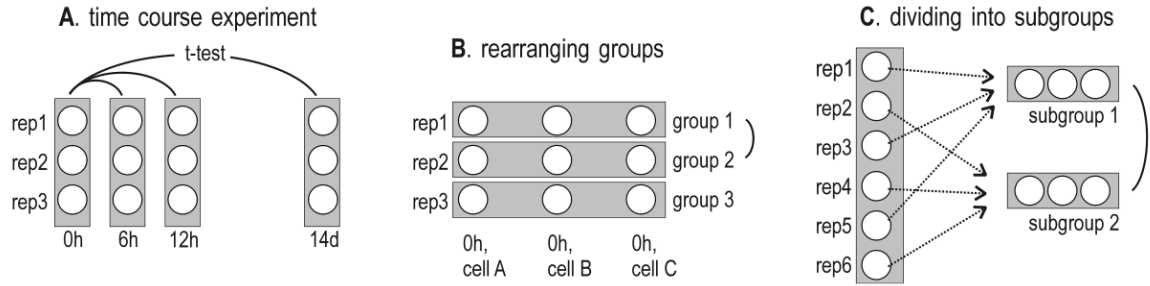

**Figure 1.** Schematic diagrams for the combinations of test subjects. The circle represents data of a sample, and the box represents a group. A is for the time-course experiment of stem cell differentiation (Hailesellasse Sene et al 2007) and its artificial imitation under true null hypotheses. The groups of time points were compared with the control 0h, in each gene of a cell line. In B, the original groups were rearranged into new groups with alternative combinations of data. Since the new groups share a population, the null hypotheses are true. However, the groups will have extra within-group variances that are derived from differences among cell lines. In C, a group with six repeats (Boutros et al 2009) was divided into two subgroups and compared. Alternative combinations were also tested. Since the data can be regarded as random samples from a population, the null hypotheses should be true.

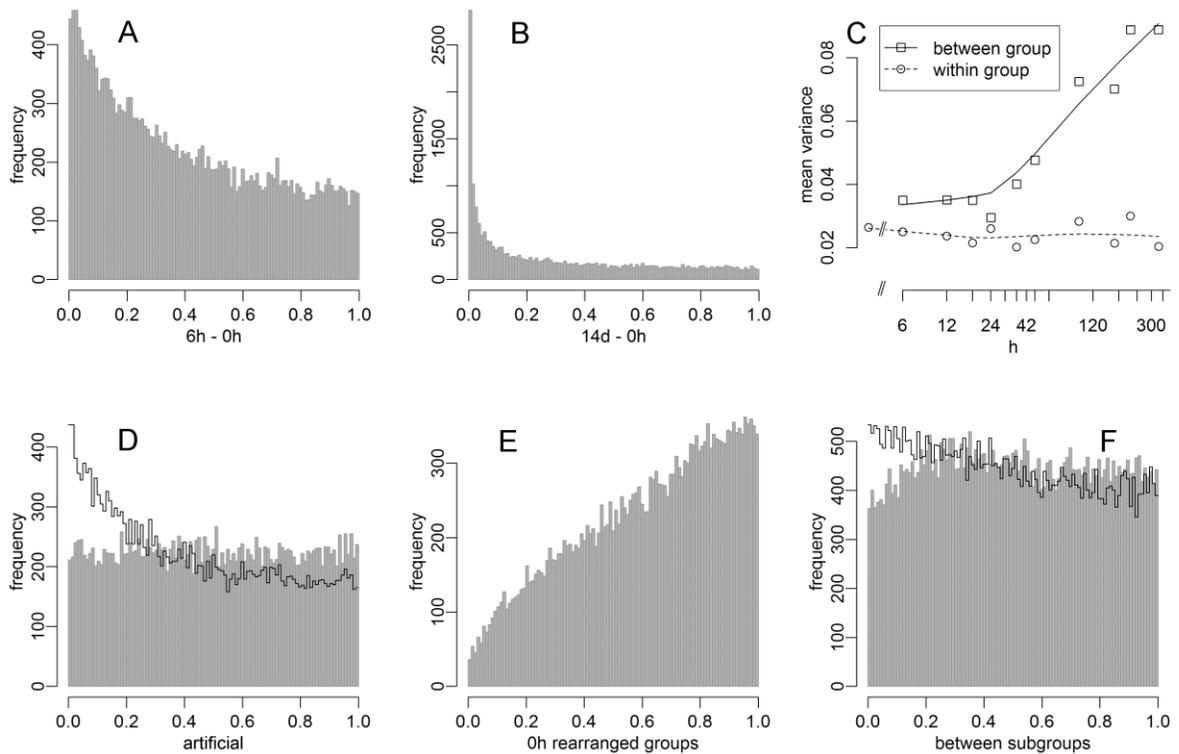

**Figure 2.** Distributions of *p*-values. Panels A-C were observed in a time-course experiment of stem cell differentiation (Hailesellasse Sene et al 2007). Panels A and B are histograms of *p*-values, observed in comparisons of 0h to 6h or 14d, respectively (observed in the series GSE2972). C shows the level of signal (solid line) and noise (dotted line). The signal is measured in the average between-group variances, and the noise is measured in the average within-group variances (Materials and Methods). D shows *p*-values observed in the artificial data when the null hypotheses are true (gray bars) or false (black line). E shows *p*-values observed between the rearranged groups (Figure 1B). F shows *p*-values observed between the artificial subgroups (Figure 1C). The gray bars and black line represent those observed in different combinations of data.

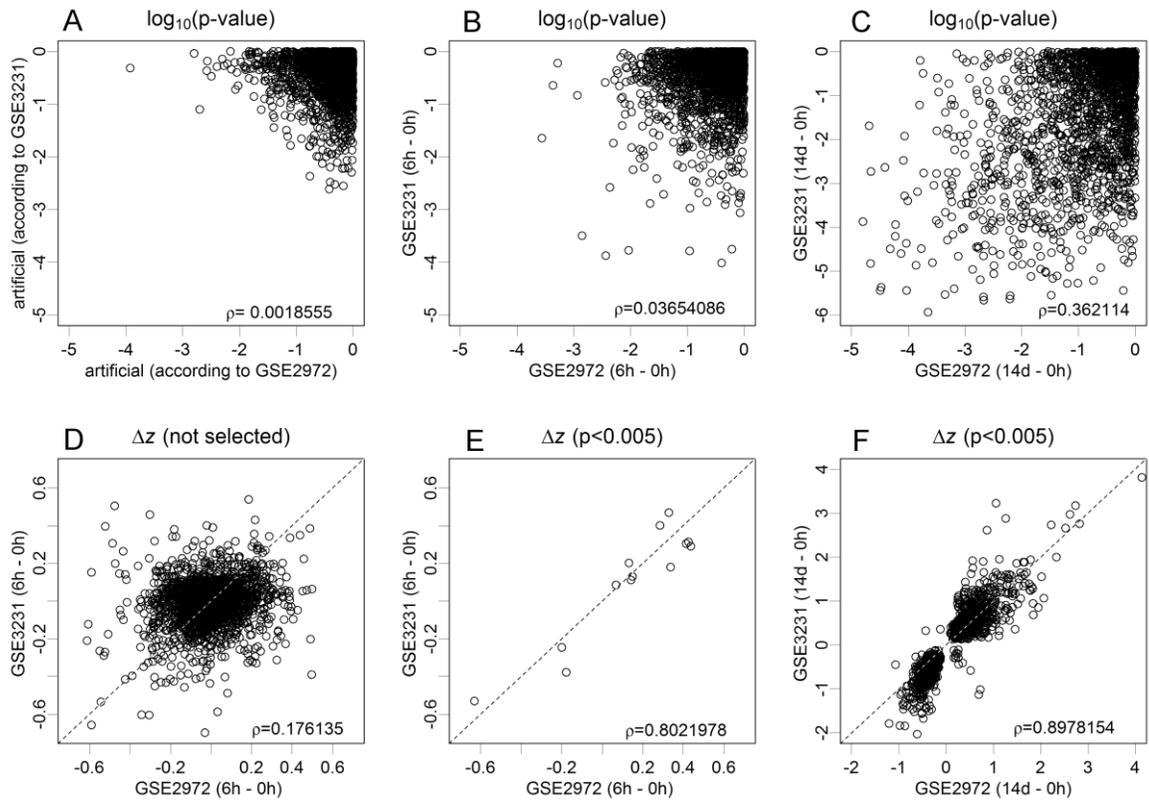

**Figure 3.** Panels A-C show relationships between *p*-values estimated in a pair of similar experiments. To observe a smaller range of *p*-values, data are presented in logarithmic axes. Panel A shows artificial data for which null hypotheses are true. B and C show microarray data, comparisons between cell lines at the indicated time points. D-F show relationships between expressional changes. D shows comparisons at 6hr, the same combination as in Panel B. Data with all the ranges of *p*-values are shown. E is similar to D, but limited to genes with *p*-values less than 0.005 in both the cell lines. F shows comparisons at 14d, in the same combination as in Panel C, genes with smaller *p*-values. Spearman's correlations are superimposed in the panels. To avoid graphical saturation, only 2,000 randomly selected genes were presented in Panels A-D.